\documentclass[amsmath,amssymb,aps,pra,preprint,showpacs,superscriptaddress]{revtex4-1}
\usepackage{graphicx}
\usepackage{dcolumn}
\usepackage{bm}
 \usepackage{color}
\begin{document}
\title{Dynamics  of different entanglement measures of two three-level atoms interacting nonlinearly with a single-mode field}
\author{H. R. Baghshahi}
\affiliation{Atomic and Molecular Group, Faculty of Physics, Yazd University, Yazd, Iran}
\affiliation{The Laboratory of Quantum Information Processing, Yazd University, Yazd, Iran}
\affiliation{Department of Physics, Faculty of Science, Vali-e-Asr University of Rafsanjan, Rafsanjan, Iran}
\author{M. K. Tavassoly}
\email[]{mktavassoly@yazd.ac.ir}
\affiliation{Atomic and Molecular Group, Faculty of Physics, Yazd University, Yazd, Iran}
\affiliation{The Laboratory of Quantum Information Processing, Yazd University, Yazd, Iran}
\date{\today}

\begin{abstract}
In this paper, we present a model which exhibits two identical $\Xi$-type three-level atoms interacting with a single-mode field with $k$-photon transition in an optical cavity enclosed by a Kerr medium. Considering full nonlinear formalism, it is assumed that the single-mode field, atom-field coupling and Kerr medium are all $f$-deformed. By using the adiabatic elimination method, it is shown that, the Hamiltonian of the considered system can be reduced to an effective Hamiltonian with two two-level atoms and $f$-deformed Stark shift. In spite of the fact that, the system seems to be complicated, under initial conditions which may be prepared for the atoms (coherent superposition of their ground and upper states) and the field (coherent state), the explicit form of the state vector of the entire system is analytically obtained. Then, the entanglement dynamics between different subsystems (i.e. ``field-two atoms'', ``atom-(field+atom)'' and ``atom-atom'') are evaluated through appropriate measures like von Neumann entropy, tangle and concurrence. In addition, the effects of intensity-dependent coupling, deformed Kerr medium, detuning parameter, deformed Stark shift and multi-photon process on the considered entanglement measures are numerically analyzed, in detail. It is shown that the degree of entanglement between subsystems can be controlled by selecting the evolved parameters, suitably. Briefly, the Kerr medium highly decreases the amount of different considered measures of entanglement, especially for two-photon transition. This destructive effect preserves even when all other parameters are present, too. Furthermore, we find that the so-called entanglement sudden death and birth
can occur in the atom-atom entanglement.
\end{abstract}
\maketitle
  \section{Introduction}\label{sec-intro}
  Entanglement is an important aspect of quantum systems which demonstrates correlations that cannot be discussed classically \cite{Nielsen}. This quantity plays a central role in many fields of research such as quantum computation \cite{Cirac1}, quantum information processing \cite{Bennett}, quantum dense coding \cite{Li1}, sensitive measurements \cite{Richter}, quantum telecloning \cite{Murao} and entanglement swapping \cite{Hu1}. Various theoretical efforts have been devoted to the classification and quantification of the entanglement. Although, the existence of entanglement between subsystems has experimentally been illustrated in different optical setups.
For instance, the atom-field entangled states have been generated through a single atom interacting with a mesoscopic field in a high-Q microwave cavity \cite{Auffeves}. In addition, quantum entanglement has been generated with transmitting two atoms in the cavity at the same time \cite{Zheng} or two atoms interacting successively with the cavity field \cite{PhoenixG}.
In this respect, a lot of schemes consist of trapped ions \cite{Turchette}, quantum dots \cite{Loss}, cavity QED \cite{Rauschenbeutel}, crystal lattices \cite{Yamaguchi}, atom-photon entanglement \cite{Phoenix} and so on have been proposed to generate the entanglement. It may be noted that the latter case is theoretically proposed by the well-known Jaynes-Cummings model (JCM) \cite{Rendell,Akhtarshenas1,Akhtarshenas2,Ouyang,Tan}.\\
The JCM \cite{Jaynes} describes the interaction between light and matter, is the simplest fully quantum-mechanical model which gives a formalism for a two-level atom interacting with a quantized radiation field in the rotating wave approximation (RWA).
In order to expand and modify this model, many generalizations such as intensity-dependent coupling \cite{Buck,Buzek,Sivakumar}, multi-level atom \cite{Yoo,Cardimona,Baghshahi2,BaghshahiQIP}, multi-photon transition \cite{Shore,Kang}, multi-atom interaction \cite{Mahmood,Baghshai1}, multi-mode field \cite{Faghihi laser,Faghihi PhyA,BaghshahiLaser}, Kerr nonlinearity \cite{Joshi,BaghshahiIJTP} etc have been reported in recent decades.
For instance, the intensity-dependent JCM was first suggested by Buck and Sukumar \cite{Buck,Sukumar} and was then used by others \cite{Liu,Yadollahi,Hekmatara,Recamier}. In detail, dynamical behavior of the JCM beyond the RWA has been studied in \cite{Naderi2011} in which the effect of the counter-rotating terms on some of the nonclassicality features has been examined. Nonclassical properties of a $\Lambda$-type three-level atom interacting with a single-mode field in an optical cavity with intensity-dependent coupling including a medium with Kerr nonlinearity in the presence of the detuning parameters have been studied by one of us \cite{Faghihi1}. More recently, by considering the full nonlinear  interaction between a two-level atom with a single-mode field, atomic population inversion as well as entropy squeezing of $k$-photon JCM in the presence of intensity-dependent Stark shift and deformed Kerr medium including time-varying atom-field coupling has been discussed by us \cite{Baghshahi}. As a more notability of the nonlinear JCM, it is instructive to state that, the ability of the nonlinear JCM in generating a class of $SU(1,1)$  coherent states of the Gilmore-Perelomov type and also $SU(2)$ group was shown by one of us \cite{Miry.Tavassoly2012}. In addition, as a results of a system in which a two-level atom interacts alternatively with a dispersive quantized cavity field and a resonant classical field, a theoretical scheme from which the nonlinear elliptical states can be generated is recently proposed \cite{Miry.etal2013}. \\
From another point of view and in direct relation to the present work, it is shown that, based on the JCM (and of course its generalization), atom-field entangled state may be generated. For instance, temporal behaviour of the von Neumann entropy of the nonlinear interaction between a $\Lambda$-type three-level atom and a two-mode cavity field in the presence of  cross-Kerr medium and its deformed counterpart \cite{Honarasa} and detuning parameters has been studied in \cite{Faghihi2,Faghihi3}. As another case, the amount of the degree of entanglement (DEM) between a three-level atom in motion interacting with a single-mode field in the intensity-dependent coupling regime is evaluated in \cite{Faghihi4}. The effects of mean photon number, detuning parameter, Kerr-like medium and various atom-field couplings on the DEM of the interaction between a $\Lambda$-type three-level atom with a two-mode field have been studied in \cite{Hanoura}.\\
From another perspective of this field of research, multi-photon JCM may be considered. Multi-photon process is of great attention in atomic systems since it leads to high degree of correlation between emitted photons resulting in the nonclassical behavior of emitted light \cite{Rodolph,Zaheer}. The importance of multi-photon transition becomes more visible when the Stark shift is taken into account. For instance, in two-photon JCM, a single-mode cavity field interacts with a two-level atom through an intermediate state containing emission or absorption \cite{Sukumar}.
It worth to be noted that, when the two atomic levels are coupled with comparable strength to the intermediate level, the Stark shift becomes considerable and cannot be ignored \cite{Alsingh,Puri1}. In this respect, by applying the adiabatic elimination method for the intermediate level(s) of a multi-level atom (such as three- or four-level one), the effective Hamiltonian containing a two-level atom together with the Stark shift is obtained \cite{Li2,Swain}. On the other hand, under particular circumstances, a multi-level atom interacting with a single-mode or multi-mode quantized field can be identified as a two-level system with Stark shift through the adiabatic elimination method \cite{Brune,Deberky,Baghshahi}. We would like to recall that, the Stark shift is usually appeared in, at least, two-photon transition \cite{Alsingh,Puri1}. \\
In this paper, we are going to outline the interaction between two three-level atoms (considering in the $\Xi$-type configuration) and a single-mode field with multi-photon transition in the presence of Kerr medium and detuning parameter. Taking nonlinear coherent states formalism into account, the single-mode field, atom-field coupling and Kerr medium are considered to be $f$-deformed. After considering all existing interactions appropriately in the Hamiltonian model of the entire atom-field system, it is found that, by applying the adiabatic elimination method, the introduced model can be reduced to a system with  effective Hamiltonian containing two two-level atoms and Stark shift effect. Then, the explicit form of the state vector of the whole system is exactly obtained by the time-dependent Schr\"{o}dinger equation which clearly possesses the entanglement properties. So, briefly speaking, the main goal of this paper is to study  individually or simultaneously the effects of deformed Stark shift, intensity-dependent coupling, deformed Kerr nonlinearity, detuning parameter and multi-photon process on the dynamics of entanglement between subsystems. For this purpose, the amount of the DEM between subsystems by evaluating von Neumann reduced entropy, tangle and concurrence are presented numerically.\\
 \section{Physical model}
 In quantum mechanics, the wave function of a physical system shows all possible information about the system. So, in the first step to analyse the system, the explicit form of the state vector of the system should be obtained. To achieve this purpose, it is necessarily required to understand all interactions which affect on the system through which one can perform the appropriate Hamiltonian. Then, with respect to the obtained Hamiltonian and also considering the (time-dependent) Schr\"{o}dinger equation or other suitable approaches \cite{Scully}, the dynamical behaviour of the state vector of the whole system can be determined. So, let us consider a model consist a single-mode field oscillating with frequency $\Omega$ which interacts with two $\Xi$-type three-level atoms in the optical cavity surrounded by a Kerr medium in the presence of detuning parameter (see figure 1). In the considered atomic system, the atomic levels are indicate as $|e\rangle$, $|i\rangle$ and $|g\rangle$ with energies $\omega_{e}$, $\omega_{i}$ and $\omega_{g}$ $(\omega_{e}>\omega_{i}>\omega_{g})$, respectively. Also, only the transitions $|e\rangle \leftrightarrow |i\rangle$ and $|i\rangle \leftrightarrow |g\rangle$ are allowed in the electric-dipole approximation \cite{Scully}. So, the Hamiltonian containing all existing interactions that describe dynamics of the considered quantum system in the rotating wave approximation can be written as
 \begin{eqnarray}\label{3}
\hat{H}&=&\Omega \hat{A}^\dagger \hat{A}+\sum_{j=1}^{2}(\omega_{e}\hat{\sigma}_{ee}^{(j)}+\omega_{i}\hat{\sigma}_{ii}^{(j)}+\omega_{g}\hat{\sigma}_{gg}^{(j)})+
 g_{1}\sum_{j=1}^{2} \left(\hat{A}^{k}\hat{\sigma}_{ei}^{(j)}+\hat{A}^{\dagger k}\hat{\sigma}_{ie}^{(j)}\right) \nonumber\\&+&g_{2}\sum_{j=1}^{2} \left(\hat{A}^{k}\hat{\sigma}_{ig}^{(j)}+\hat{A}^{\dagger k}\hat{\sigma}_{gi}^{(j)}\right) +\chi \hat{A}^{\dagger 2} \hat{A}^{2},
\end{eqnarray}
 where $g_{i}$ ($i=1,2$) refers to the coupling constants, $k$ indicates multi-photon process, $\chi$ shows dispersive part of
the third-order nonlinearity of the Kerr medium and $\hat{\sigma}_{ij}$ denotes the atomic ladder operator which is defined by $\hat{\sigma}_{ij}=|i\rangle \langle j|$ ($i,j=e,i,g$). Also, the operators $\hat{A}=\hat{a}f(\hat{n})$ and $\hat{A}^{\dagger}=f(\hat{n})\hat{a}^{\dagger}$ with $\hat{n}=\hat{a}^{\dagger}\hat{a}$ are respectively the nonlinear ($f$-deformed) annihilation and creation operators, which satisfy the following communication relations:
 \begin{eqnarray}\label{4}
[\hat{A},\hat{A}^{\dagger}]=(n+1)f^{2}(n+1)-nf^{2}(n),\hspace{0.5cm}
[\hat{A},\hat{n}]=\hat{A},\hspace{0.5cm} [\hat{A}^{\dagger},\hat{n}]=-\hat{A}^{\dagger}.
\end{eqnarray}
\begin{equation}
\left(\begin{array}{cc|cc}
\rho_{11}&\rho_{12}&\rho_{13}&\rho_{14}  \\
\rho_{21}&\rho_{22}&\rho_{23}&\rho_{24}\\\hline
\rho_{31}&\rho_{32}&\rho_{33}&\rho_{34}\\
\rho_{41}&\rho_{42}&\rho_{43}&\rho_{44}
\end{array}\right)\quad\rightarrow
\left(\begin{array}{cccc}
\rho_{11}&\rho_{21}&\rho_{12}&\rho_{22}\\\hline
\rho_{31}&\rho_{41}&\rho_{32}&\rho_{42}\\\hline
\rho_{13}&\rho_{23}&\rho_{14}&\rho_{24}\\\hline
\rho_{33}&\rho_{43}&\rho_{34}&\rho_{44}
\end{array}\right)
\end{equation}

  In these relations, $f(\hat{n})$ is a Hermitian operator-valued function responsible for the intensity-dependent function. To proceed further, we define the detuning parameters in the form $\Delta_{1}=k\Omega-(\omega_{i}-\omega_{g})$ and $\Delta_{2}=k\Omega-(\omega_{e}-\omega_{i})$, and assume that $\Delta_{1}=-\Delta_{2}=\delta$ (the adiabatic elimination condition).
     Now recalling the adiabatic elimination method \cite{Puri1,Ahmad}, the effective Hamiltonian describing the system  reads as follows:
 \begin{eqnarray}\label{41}
\hat{H}&=&\Omega \hat{A}^\dagger \hat{A}+\frac{1}{2}\omega\sum_{j=1}^{2}\sigma_{z}^{(j)}+
 g\sum_{j=1}^{2} \left(\hat{A}^{2k}\hat{\sigma}_{eg}^{(j)}+\hat{A}^{\dagger 2k}\hat{\sigma}_{ge}^{(j)}\right) \nonumber\\&+&\hat{A}^{\dagger k} \hat{A}^{k}\sum_{j=1}^{2}\left(\beta_{1} \sigma_{ee}^{(j)}+\beta_{2} \sigma_{gg}^{(j)}\right)+\chi \hat{A}^{\dagger 2} \hat{A}^{2},
\end{eqnarray}
 where $\beta_{1}=2g_{1}^{2}/ \delta$, $\beta_{2}=2g_{2}^{2}/ \delta$ correspond to the parameters related to Stark shift and $g=\sqrt{\beta_{1}\beta_{2}}$, $\sigma_{z}^{(j)}=\sigma_{ee}^{(j)}-\sigma_{gg}^{(j)}$ and $\omega=\omega_{e}-\omega_{g}$.
 It is obvious that the above Hamiltonian in the absence of Stark shift and Kerr medium corresponds to the generalized JCM for $2k$-photon transitions. By considering $\Delta=\omega-2k\Omega$, the Hamiltonian (3) can be rewritten in the following form
 \begin{eqnarray}\label{5}
\hat{H}_{\mathrm{eff}}&=&\Omega \Big(\hat{A}^{\dag} \hat{A}+k\sum_{j=1}^{2}\hat{\sigma}_{z}^{(j)}\Big)+\frac{\Delta}{2}\sum_{j=1}^{2}\hat{\sigma}_{z}^{(j)}+ g\sum_{j=1}^{2} \left(\hat{A}^{2k}\hat{\sigma}_{eg}^{(j)}+\hat{A}^{\dagger 2k}\hat{\sigma}_{ge}^{(j)}\right) \nonumber\\&+&\hat{A}^{\dagger k} \hat{A}^{k}\sum_{j=1}^{2}\left(\beta_{1} \sigma_{ee}^{(j)}+\beta_{2} \sigma_{gg}^{(j)}\right)+\chi \hat{A}^{\dagger 2} \hat{A}^{2}.
\end{eqnarray}
 It is instructive to emphasis on the fact that, the fourth term in the above Hamiltonian shows the $f$-deformed Stark shift which means that the energy shifts of the atomic levels $|e\rangle$ and $|g\rangle$ depend on the intensity of light.
Anyway, according to the Hamiltonian in (\ref{5}), we are dealing with the effective Hamiltonian which describes two two-level atoms interacting with a single-mode field in the presence of $f$-deformed Kerr medium and intensity-dependent coupling with multi-photon process and also `nonlinear' Stark shift.
Now, keeping in mind all terms of the Hamiltonian in (\ref{5}) together with paying careful attention to the allowable transitions, one may consider the wave function $|\psi(t)\rangle$ at any time $t>0$ in the form
\begin{eqnarray}\label{8}
|\psi(t)\rangle&=&\sum_{n=0}^{\infty} \exp \left [-i\omega t \left (\hat{n}f^{2}(\hat{n}) +  k \sum_{j=1}^{2} \hat{\sigma}_{z}^{(j)} \right) \right]
\times
\bigg [ A(n,t) e^{-i\gamma_{1}t}|e,e,n\rangle\nonumber\\&+&B(n+2k,t)e^{-i\gamma_{2}t}(|e,g,n+2k\rangle+|g,e,n+2k\rangle)\nonumber\\&
+&C(n+4k,t)e^{-i\gamma_{3}t}|g,g,n+4k\rangle \bigg ],
\end{eqnarray}
where $A$, $B$ and $C$ are the atomic probability amplitudes which have to be evaluated, and
 \begin{eqnarray}\label{9}
\gamma_{1}&=&\Delta+ \chi n (n-1) f^{2}(n) f^{2}(n-1)+ 2 \beta_{1} \frac{n!}{(n-k)!}  \left (\frac{[f(n)]!}{[f(n-k)]!}\right) ^{2} ,\nonumber\\
\gamma_{2}&=&\chi (n+2k)(n+2k-1) f^{2}(n+2k)f^{2}(n+2k-1) \nonumber\\&+&(\beta_{1}+\beta_{2}) \frac{(n+2k)!}{(n+k)!} \left(\frac{[f(n+2k)]!}{[f(n+k)]!}\right)^{2},\nonumber\\
\gamma_{3}&=&-\Delta+\chi (n+4k)(n+4k-1)f^{2}(n+4k)f^{2}(n+4k-1) \nonumber\\&+& 2 \beta_{2}\frac{(n+4k)!}{(n+3k)!} \left(\frac{[f(n+4k)]!}{[f(n+3k)]!}\right)^{2},
\end{eqnarray}
where $[f(n)]!=f(n)f(n-1)...f(1)$.
Now, by substituting $|\psi(t)\rangle$ from equation (\ref{8}) and $\hat{H}_{\mathrm{eff}}$ from equation (\ref{5})  in the time-dependent Schr\"{o}dinger equation ($i\frac{\partial}{\partial t}|\psi(t)\rangle=\hat{H}_{\mathrm{eff}}|\psi(t)\rangle$), one may arrive at the following coupled differential equations for the atomic probability amplitudes
\begin{eqnarray}\label{10}
i \dot{A}&=& 2 V_{1} B e^{-i\eta t},\nonumber\\
i \dot{B}&=& V_{1} A e^{i\eta t}+V_{2} C e^{-i\varsigma t},\nonumber\\
i \dot{C}&=& 2 V_{2} B e^{i\varsigma t},
\end{eqnarray}
where the dot sign refers to the time differentiation and we have defined
\begin{eqnarray}\label{11}
  V_{1}&=& g \sqrt{\frac{(n+2k)!}{n!}}\frac{[f(n+2k)]!}{[f(n)]!},\hspace{0.5cm}  V_{2}= g \sqrt{\frac{(n+4k)!}{(n+2k)!}}\frac{[f(n+4k)]!}{[f(n+2k)]!},\nonumber\\
  \eta&=&\gamma_{2}-\gamma_{1}, \hspace{0.5cm} \varsigma=\gamma_{3}-\gamma_{2}.
\end{eqnarray}
By setting $C=e^{i \mu t}$ and substituting it into equation (\ref{10})  the following third order algebraic equation may be obtained
\begin{equation}\label{12}
\mu^{3}+x_{1}\mu^{2}+x_{2}\mu+x_{3}=0,
\end{equation}
where
\begin{equation} \label{13}
x_{1}=-\eta-2\varsigma, \hspace{0.5cm}x_{2}=\varsigma(\varsigma+\eta)-2(V_{1}^{2}+V_{2}^{2}), \hspace{0.5cm}x_{3}=2V_{2}^{2}(\eta+\varsigma).
  \end{equation}
  Equation (\ref{12}) has generally three different roots which are given by \cite{Childs}
 \begin{eqnarray} \label{14}
 \mu_{m}&=&-\frac{1}{3}x_{1}+\frac{2}{3}\sqrt{x_{1}^{2}-3x_{2}}\cos \left(\phi+\frac{2}{3}(m-1)\pi\right), \hspace{0.5cm}m=1,2,3,
\nonumber\\ \phi&=&\frac{1}{3}\cos^{-1}\left[ \frac{9 x_{1}x_{2}-2x_{1}^{3}-27x_{3}}{2(x_{1}^{2}-3x_{2})^{3/2}}\right].
  \end{eqnarray}
  Consequently, coefficient $C$ should be considered as a linear combination of $e^{i \mu_{m} t}$, that is,
  \begin{equation} \label{15}
  C=\sum_{m=0}^{3}b_{m}e^{i \mu_{m} t}.
   \end{equation}
   Now, by inserting (\ref{15}) into (\ref{10}) and after straightforward calculations one may find the probability amplitudes in the forms:
 \begin{eqnarray} \label{16}
 &A&(n,t)=\frac{1}{2V_{1}V_{2}}\sum_{m=0}^{3}b_{m} (\mu_{m}^{2}-\varsigma\mu_{m}-2V_{2}^{2})e^{i(\mu_{m}-\eta-\varsigma)t},\nonumber\\
 &B&(n+2k,t)=\frac{-1}{2V_{2}}\sum_{m=0}^{3}\mu_{m}b_{m}e^{i(\mu_{m}-\varsigma)t},\nonumber\\
 &C&(n+4k,t)=\sum_{m=0}^{3}b_{m}e^{i\mu_{m}t},
  \end{eqnarray}
  where $b_{m}$ may be evaluated by applying the initial conditions for the atoms and field. To reach this goal, suppose that the atoms enter the cavity in the coherent superposition of the exited  ($|e,e\rangle$) and  ground ($|g,g\rangle$) states, i.e.,
\begin{equation} \label{18}
|\Psi_{atoms}(t=0)\rangle=\cos(\theta/2)|e,e\rangle+\sin(\theta/2)|g,g\rangle.
\end{equation}
  It is worthwhile to mention that, arbitrary amplitudes of the initial state of the field such as thermal, number, coherent or squeezed state can be considered. However, because of the fact that coherent state (the laser field far above the threshold condition \cite{Scully}) is more accessible than other typical field states, we assume that the field initially to be in a standard coherent state with mean photon number $|\alpha|^{2}$ as
\begin{equation} \label{17}
|\alpha\rangle=\sum_{n=0}^{\infty}q_{n}|n\rangle,\hspace{0.5cm}q_{n}=\exp(-\frac{|\alpha|^{2}}{2})\frac{\alpha^{n}}{\sqrt{n!}}.
\end{equation}
By applying these initial conditions for atoms and field and using equation (\ref{16}), the $b_{m}$ coefficients read as
  \begin{equation} \label{20}
b_{m}=\frac{2V_{1}V_{2}q_{n}\cos(\theta/2)+(2V_{2}^2+\mu_{k}\mu_{l}) q_{n+4k} \sin(\theta/2)}{\mu_{mk}\mu_{ml}},\hspace{2cm}m\neq k \neq l=1,2,3,
\end{equation}
where $\mu_{mk}=\mu_{m}-\mu_{k}$. Consequently, the probability amplitudes $A$, $B$ and $C$ are precisely derived.
At last, it should necessarily be noticed that, choosing different nonlinearity functions clearly leads to different Hamiltonian systems and so different physical results may be achieved. In the present model, we select the particular deformation function $f(n) = \sqrt{n}$. This function is a favorite function for the authors who have worked in the nonlinear regimes of atom-field interaction (see for instance \cite{Agarwal,Huang}). As a physical motivation of choosing this nonlinearity function, it is valuable to state that, Fink \textit{et al} have discovered a physical system in which this special nonlinearity function is naturally appeared \cite{Fink}.
Anyway, in the next section, the DEM between different subsystems is numerically examined by evaluating suitable entanglement measures.
\section{Quantum entanglement}
The entanglement connected to a multipartite system indicates the joint information between the subsystems. The DEM between subsystems  corresponds to the nonlocal information in the system, so that the higher order of DEM is proportional to less information about the subsystems \cite{Bennett25}. Since the atom-field interaction is usually regarded as a simple way to produce the entangled states, it is valuable to examine the amount of entanglement between subsystems (here the two atoms and the field as a tripartite system).
To achieve the latter purpose, as a few measures leading to the amount of the DEM we may refer to, for instance, von Neumann entropy and relative entropy \cite{Plenio}, linear entropy \cite{Kim},  entanglement of formation \cite{Wootters}, concurrence \cite{Hill}, negativity \cite{Peres} and tangle \cite{Osborne}. So, in this section, the amount of DEM between different subsystems such as atom-field through von Neumann reduced entropy, ``atom$+$field''-atom via tangle and atom-atom by concurrence will be evaluated numerically.\\
For evaluating these entanglement measures, the reduced density matrix of the two atoms (for von Neumann reduced entropy and concurrence) and the reduced density matrix for one of the atoms (for tangle) are required. At any time $t>0$ the reduced density matrix of the two atoms for the considered system is given by:
\begin{equation} \label{1717}
\hat{\rho}_{A}(t)=\mathrm{Tr}_{F}(\hat{\rho}_{AF}(t))=\mathrm{Tr}_{F}(|\psi(t)\rangle \langle \psi(t)|)=
\left(
  \begin{array}{cccc}
    \rho_{11}(t) & \rho_{12}(t)  & \rho_{13}(t)  & \rho_{14}(t)  \\
   \rho_{21}(t) & \rho_{22}(t)  & \rho_{23}(t)  & \rho_{24}(t) \\
   \rho_{31}(t) & \rho_{32}(t)  & \rho_{33}(t)  & \rho_{34}(t)\\
   \rho_{41}(t) & \rho_{42}(t)  & \rho_{43}(t)  & \rho_{44}(t) \\
  \end{array}
\right),
\end{equation} \label{1718}
with the matrix elements
 \begin{eqnarray} \label{1719}
  \rho_{11}(t)&=&\sum_{n=0}^{\infty} A(n,t)A^{\ast}(n,t),\nonumber\\
    \rho_{12}(t)&=&\rho_{13}(t)=\sum_{n=0}^{\infty} A(n+2k,t)B^{\ast}(n+2k,t) e^{i\Re_{1}},\nonumber\\
     \rho_{14}(t)&=&\sum_{n=0}^{\infty} A(n+4k,t)C^{\ast}(n+4k,t) e^{2i\Re_{2}},\nonumber\\
       \rho_{22}(t)&=&\rho_{23}(t)=\rho_{33}(t)=\sum_{n=0}^{\infty} B(n+2k,t)B^{\ast}(n+2k,t), \nonumber\\
       \rho_{24}(t)&=&\rho_{34}(t)=\sum_{n=0}^{\infty} B(n+4k,t)C^{\ast}(n+4k,t) e^{i\Re_{2}},\nonumber\\
        \rho_{44}(t)&=&\sum_{n=0}^{\infty} C(n+4k,t)C^{\ast}(n+4k,t).
 \end{eqnarray}
 where
  \begin{eqnarray} \label{17191}
  \Re_{1}&=&-\Delta+ (\beta_{2}-\beta_{1})\frac{(n+2k)!}{(n+k)!} \left(\frac{[f(n+2k)]!}{[f(n+k)]!}\right)^{2},\nonumber\\
   \Re_{2}&=&-\Delta+ (\beta_{2}-\beta_{1})\frac{(n+4k)!}{(n+3k)!} \left(\frac{[f(n+4k)]!}{[f(n+3k)]!}\right)^{2}.
   \end{eqnarray} \label{17191}
 Also, the reduced density matrix of the first atom ($A_{1}$) can be obtained by tracing equation (\ref{1717}) over the second atom ($A_{2}$) as follows:
 \begin{equation} \label{1720}
\hat{\rho}_{A_{1}}(t)=\mathrm{Tr}_{A_{2}}(\hat{\rho}_{A}(t))=
\left(
   \begin{array}{cc}
     y_{11}(t) & y_{12}(t) \\
      y_{21}(t)&  y_{22}(t) \\
   \end{array}
 \right)
  \end{equation} \label{1721}
 where $y_{11}(t)=  \rho_{11}(t) +  \rho_{22}(t)$, $y_{12}(t)=  \rho_{12} (t)+  \rho_{24}(t)$ and $y_{22}(t)=  \rho_{22} (t)+  \rho_{44}(t)$.\\
 Now, we are ready to present our numerical results and analyze them. Before every thing it should be mentioned that, in all figures which will be presented in this section (figures 2-4) the time evolution of the mentioned quantities in terms of the scaled time $gt$ for the two atoms prepared initially in the exited state and the field in coherent state with $|\alpha|^{2}=25$ and also $f(n)=\sqrt{n}$ (intensity-dependent coupling) are calculated.
Different plots of each figure have been set in two columns. The left (right) column corresponds to $k=1$  ($k=2$ ) (two- (four-) photon transition).
The details of the various plotted graphs are as below. In frame (a) the Kerr and Stark effects are absent ($\chi=0$, $\beta_{1}=0=\beta_{2}$) and the exact resonant case was assumed ($\Delta=0$). Frame (b) shows the effect of detuning parameter ($\Delta=10$) without Kerr and Stark effects ($\chi=0$, $\beta_{1}=0=\beta_{2}$). The influence of the Kerr nonlinearity ($\chi=0.5$) in the absence of detuning parameter and Stark shift ($\Delta=0$, $\beta_{1}=0=\beta_{2}$) is depicted in frame (c). Frame (d) is plotted for searching the particular effect of the Stark shift ($\beta_{1}=6$, $\beta_{2}=1$) in the absence of the Kerr medium and detuning parameter ($\chi=0$, $\Delta=0$). Finally, the influence of detuning parameter ($\Delta=10$) in the presence of Kerr medium ($\chi=0.5$) and Stark shift ($\beta_{1}=6$, $\beta_{2}=1$) is represented in frame (e).
 \subsection{The entanglement between ``the two atoms'' and ``the field'': von Neumann entropy}
It is shown that, for any multipartite quantum system that is described by a pure state, the quantum entropy is a suitable measure to obtain the DEM between subsystems \cite{Phoenix,BuzekVon}.
The considered system is made by three entities: two atoms and a single-mode field. Now, before studying the entanglement between them, it is required to pay attention to the important theorem of Araki and Leib \cite{Araki}. They showed that for a bipartite quantum system, the entropy of system and subsystems (here atoms and field) at any time $t$ are limited to the triangle inequalities $|S_{A}(t)-S_{F}(t)|\leq S_{AF}(t)\leq|S_{A}(t)+S_{F}(t)|$, where $S_{AF}$ shows the total entropy of the system and the subscripts `A' and `F' refer to the atom and field, respectively. As a consequence of this inequality, if the system starts from a pure state (as we have considered), the total entropy of the system is zero and remains constant. This means that at any time $t > 0$, the atomic and field reduced entropies are equal, that is, $S_{A}(t) = S_{F} (t)$ \cite{PhoenixVon,BarnettVon}. Consequently, we only need to evaluate the reduced entropy of the atom/field, in order to obtain DEM. According to the von Neumann entropy, the reduced entropy of the atom (field) is defined through the corresponding reduced density operators by
 \begin{equation} \label{171}
S_{A(F)}(t) = - \mathrm{Tr}_{A(F)}(\hat{\rho}_{A(F)}(t) \ln \hat{\rho}_{A(F)}(t)),
\end{equation}
where $\hat{\rho}_{A(F)}(t)=\mathrm{Tr}_{F(A)}(|\psi(t)\rangle \langle\psi(t)|)$.
It is preferable to use the basis in which the atomic density matrix is diagonal. So, the equation (\ref{171}) can be rewritten in the form
\begin{equation} \label{172}
\mathrm{DEM}(t)=S_{F}(t)=S_{A}(t)=-\sum_{j=1}^{4}\lambda_{j} \ln\lambda_{j},
\end{equation}
where $\lambda_{j}$, the eigenvalues of the reduced density matrix of the two atoms, are given by the Cardano's method as \cite{Childs}
\begin{eqnarray}\label{173}
\lambda_{j}&=&-\frac{1}{3}\xi_{1}+\frac{2}{3}\sqrt{\xi_{1}^{2}-3\xi_{2}}\cos \left(\beta+\frac{2}{3}(j-1)\pi\right),\hspace{0.5cm}j=1,2,3,\nonumber\\
\lambda_{4}&=0&
\end{eqnarray}
with
\begin{equation} \label{174}
\beta=\frac{1}{3}\cos^{-1}\left[ \frac{9 \xi_{1}\xi_{2}-2\xi_{1}^{3}-27\xi_{3}}{2(\xi_{1}^{2}-3\xi_{2})^{3/2}}\right],
\end{equation}
and
\begin{eqnarray}\label{175}
\xi_{1}&=&-\rho_{11}-2\rho_{22}-\rho_{44},\nonumber\\
\xi_{2}&=&-2\rho_{12}\rho_{21}-\rho_{14}\rho_{41}-2\rho_{24}\rho_{42}+2\rho_{22}\rho_{44}+\rho_{11}(2\rho_{22}+\rho_{44}),\nonumber \\
\xi_{3}&=&2\Bigg(\rho_{14}(\rho_{22}\rho_{41}-\rho_{21}\rho_{42})+\rho_{12}(\rho_{21}\rho_{44}-\rho_{24}\rho_{41})\nonumber\\&+&\rho_{11}(\rho_{24}\rho_{42}
-\rho_{22}\rho_{44})\Bigg),
\end{eqnarray}
where $\rho_{ij}(t)$ are the elements of the reduced density matrix  of the two atoms that have been introduced in equation (\ref{1719}).
As the first measure of DEM between the two atoms and the field,  the time evolution of the field entropy has been plotted in terms of the scaled time $gt$ in figure 2.
 Figure 2(a) shows the time variation of  this quantity without Kerr and Stark effects in the resonance condition. The oscillations of DEM with time is irregular for both cases, i.e., two- and four-photon transitions.
    As is seen, in particular,  in the case $k=1$ at some moments of time the entropy suddenly decreases to nearly zero value.   So, comparing the left and right plots of 2(a) shows that changing  the value of $k$ from 1 to 2 increases the minima of DEM. In figure 2(b) the influence of detuning parameter in the absence of Kerr nonlinearity and Stark shift is examined. Comparing 2(b) with 2(a) it appears that the presence of the detuning parameter do not  effect on the entropy of the system critically. From figure 2(c) by which the effect of Kerr nonlinearity is revealed, one can see  that the amount of entanglement is drastically decreased for two-photon transition. But the effect of this parameter in the case of four-photon transition is not so significant.
 We discuss the effect of Stark shift on the time evolution of DEM in figure 2(d). The effect of  this phenomenon on the four-photon process is negligible and it decreases the amount of DEM for two-photon transition as compared with 2(a), 2(b).  Figure 2(e) demonstrates the effects of detuning, Kerr medium and Stark shift simultaneously.
   What happens in this case is very similar to the case 2(c). Generally in all cases discussed above, two-photon transition ($k=1$) is more sensitive to the absence/presence of mentioned parameters (detuning parameter, Kerr nonlinearity and Stark shift) in comparison with four-photon case ($k=2$).
 \subsection{The entanglement between ``one atom'' and ``the reminder of the system'': tangle}
 Tangle is a good measurement of entanglement which describes the correlation between subsystems of a multipartite quantum system. At first, this measure was defined in terms of concurrence for two qubits \cite{Coffman}. This definition was extended to obtain an analytical form of the tangle for a bipartite system in a pure state $|\psi_{AB}\rangle$ with arbitrary dimensions $D_{1} \times D_{2}$, which is given by \cite{Rungta,Tessier}
\begin{equation} \label{T1}
\tau(\psi_{AB})=2 \nu_{D_{_{{1}}}} \nu_{D_{_{{2}}}}\left[1- \mathrm{Tr}(\rho_{A}^{2})\right],
\end{equation}
where $\nu_{D_{_{{1}}}}$ and $\nu_{D_{_{{2}}}}$ are arbitrary factors, which may in general depend on the dimensions of the subsystems $D_{1}$ and $D_{2}$, respectively. A sensible selection for the constants $\nu_{D_{_{{1}}}}$ and $\nu_{D_{_{{2}}}}$ is the value one, in order to be in agreement with the two-qubit case. So with this choice, tangle runs from $0$ for separable (product) state to $2 (M-1)/M$ for a maximally entangled state with $M=\min(D_{1},D_{2})$.
Any way, because of  the symmetry in the considered system (two identical two-level atoms and a single-mode cavity field), we examined the entanglement between one atom and the reminder of the system by using of tangle measure. By considering $A_{1}$, $A_{2}$ and $F$ as  labels of the first atom, second atom and the field, respectively, the tangle reads as
  \begin{eqnarray} \label{T2}
\tau_{A_{1}-A_{2}F}&=&2\left[1- \mathrm{Tr}(\rho_{A_{1}}^{2})\right],
\end{eqnarray}
where  $\tau_{A-B}$ shows the DEM between bipartite $A$ and $B$ and $\rho_{j}$ is the reduced density matrix for the \textit{j}th party, which can be obtained by using of density matrix for the whole system (equation (\ref{1720})). It ought to be mentioned that in the case defined in (\ref{T2}), the tangle is limited to the inequality $0\leq\tau_{A_{1}-A_{2}F}\leq1$.\\
 We have plotted figure 3 for studying the DEM between the first atom and the reminder of the system ($\tau_{A_{1}-A_{2}F}$) by considering the same chosen values have been used in figure 2. As may be seen the behaviour of this measure is qualitatively the same as figure 2, even though not exactly quantitatively.
   However, as  is expected the upper bound of this criterion which measures the entanglement between the first atom with the second atom and the field is 1.
 \subsection{The entanglement between the two atoms: concurrence}
The main goal of this subsection is to investigate the DEM between two atoms that the corresponding state is mixed. It is valuable to
mention that, two-atom entangled states have experimentally been reported by considering ultra cold trapped ions \cite{DeVoe1996} and cavity QED schemes \cite{Hagley1997}.
Concurrence and negativity are the suitable measures for studying the DEM between the atoms \cite{HorodeckiMod}.
Here, we examine the temporal behaviour of the concurrence.
This quantity for a pair of qubits ($A$ and $B$) with density matrix $\rho_{AB}$ that can be mixed or pure, is given by  \cite{Wootters,Coffman}
 \begin{equation} \label{C1}
\mathcal{C}_{AB}=\max \left \{0,2 \max [\lambda_{j}]- \sum_{j=1}^{4} \lambda_{j} \right\},
\end{equation}
where $\lambda_{j}$ are the square roots of the eigenvalues of the operator $\tilde{\rho}_{AB}\rho_{AB}$, in which $\tilde{\rho}_{AB}$ showing the  ``spin-flipped'' density matrix, which is defined by
 \begin{equation} \label{C2}
\tilde{\rho}_{AB}=(\sigma_{y} \otimes \sigma_{y}) \rho_{AB}^{\ast} (\sigma_{y} \otimes \sigma_{y}),
\end{equation}
with $\rho_{AB}^{\ast}$ as the complex conjugate of $\rho_{AB}$.
Since, we are going to study the dynamics of entanglement between the two atoms, in the present formalism, $\rho_{AB}$ replaced by $\rho_{A_{1}A_{2}}$ exhibiting the reduced density operator of the atoms (equation (\ref{1717})).\\
In figure 4, we present our numerical results of the concurrence for measuring the entanglement between the two atoms for different chosen parameter similar to  figures 2 and 3.
Comparing all plots in this figure it is seen that, in the absence of all parameters the difference between $k=1$, $k=2$ are negligible.
Adding detuning to both cases decreases the concurrence as time goes on, however, this will be faster for the case of four-photon transition (figure 4(b)).
The effect of Kerr medium is shown in  4(c). This nonlinearity causes a critical  decrease in the value of concurrence, especially for the two-photon transition.
 While the effect of Stark shift is again a decrease in the considered criterion, in contrast to the latter case, the decrease will be higher for the four-photon case (figure 4(d)). Figure 4(e) is similar to 4(c). Also, the entanglement sudden death and birth \cite{Laurat,Hu} can be seen in figures 4(b)-4(d). Particularly, these phenomena increases in the presence of Stark shift and Kerr medium.
\section{Summary and conclusion}
In this paper we have studied two $\Xi$-type three-level atoms interacting with a single-mode field with multi-photon process in an optical cavity containing a Kerr medium in the presence of intensity-dependent coupling and detuning parameter. By applying the adiabatic elimination method, we found that, the Hamiltonian of the system can be reduced to the effective Hamiltonian containing two two-level atoms and Stark shift effect. Next, after finding the explicit form of the state vector of the whole system analytically, the effects of intensity-dependent coupling, detuning parameter, Kerr nonlinearity, multi-photon transition and Stark shift on the temporal behaviour of the well-known entanglement criteria have been numerically investigated. For this purpose, the time evolution of the von Neumann reduced entropy (for studying the entanglement between the atoms and field), the tangle (for investigating the DEM between one of the atoms and the other subsystems) and concurrence (for evaluating the atom-atom entanglement) have been examined, in detail. Briefly, the main results of the paper are as follow.\\
 Summing up, the amount of entanglement between subsystems can be controlled appropriately through entering the considered physical parameters.
 The numerical results related to the field entropy and tangle (figures 2 and 3) show that for $k=2$ (four-photon transition), the presence or absence of all or a few of the considered effects (detuning parameter, Kerr medium and Stark shift) have no significant role on the amount and behavior of the entanglement measures. In these cases, it is found that, the amount of DEM remains around the maximum mean value when the time proceeds. In the case $k=1$, the maximum value of DEM is attained when all parameters are absent. Also, the existence of the detuning or Stark effect may causes a little reduction in the DEM, however, in the presence of Kerr nonlinearity, as well as all three parameters containing the kerr effect, the DEM is drastically decreased.\\
Different features of the temporal behavior of concurrence (figure 4) may be observed when the related  numerical results are compared with two previous measures (tangle and entropy). It is shown that, for both  $k=1$ and $k=2$, the absence of all parameters leads to the maximally atom-atom entangled state. Furthermore, the existence of considered parameters may individually and simultaneously reduce the DEM between the atoms. In these cases the entanglement sudden death and birth have been observed.
In detail, the numerical results of concurrence for $k=1$ indicate that, the presence of Kerr medium and also simultaneous effects of Kerr medium, Stark shift and detuning parameter decrease the amount of DEM, considerably. It may be noted that, the reduction of the DEM for $k = 2$ is less than for the case $k=1$.
Moreover, with constant coupling ($f(n)=1$) and by using the parameters in Refs. \cite{Ahmad,Hosny} our numerical results successfully recover the results in Refs. \cite{Ahmad,Hosny} in the case of two-photon transitions ($k=1$).
 \begin{acknowledgements}
The authors would like to thank the referee for his/her helpful comments and suggestions which  satisfactorily improved the contents of the paper.
 \end{acknowledgements}
\section*{References}
\providecommand{\newblock}{}

\end{document}